\documentclass[preprint]{aastex62}
\usepackage{array}

\usepackage{color}

\newcommand{\CC}{{\tt C1.0+/0/24}}
\newcommand{\MM}{{\tt M1.0+/0/24}}
\newcommand{\Cp}{{\tt C1.0+}}
\newcommand{\Mp}{{\tt M1.0+}}
\newcommand{\Mfp}{{\tt M5.0+}}
\newcommand{\Xp}{{\tt X1.0+}}
\newcommand{\Xfp}{{\tt X5.0+}}
\newcommand{\Co}{{\tt C1.0--C9.9}}
\newcommand{\Mo}{{\tt M1.0--M9.9}}

\received{}
\revised{}
\accepted{}

\submitjournal{ApJ}

\begin{document}

%\title{The Performance Characteristics of Operational Flare Forecasting Systems II: 
%Systematic Behaviors}
\title{A Comparison of Flare Forecasting Methods. III. Systematic Behaviors of Operational Solar Flare Forecasting Systems}

\correspondingauthor{K.~D.~Leka}
\email{kdleka@isee.nagoya-u.ac.jp, leka@nwra.com}

\author[0000-0003-0026-931X]{K.D. Leka}
\affiliation{Institute for Space-Earth Environmental Research
Nagoya University 
Furo-cho Chikusa-ku 
Nagoya, Aichi 464-8601 JAPAN}
\affiliation{NorthWest Research Associates
3380 Mitchell Lane
Boulder, CO 80301 USA}

\author[0000-0001-9149-6547]{Sung-Hong Park}
\affiliation{Institute for Space-Earth Environmental Research
Nagoya University 
Furo-cho Chikusa-ku 
Nagoya, Aichi 464-8601 JAPAN}

\author[0000-0002-6814-6810]{Kanya Kusano}
\affiliation{Institute for Space-Earth Environmental Research
Nagoya University 
Furo-cho Chikusa-ku 
Nagoya, Aichi 464-8601 JAPAN}

\author{Jesse Andries}
\affiliation{STCE - Royal Observatory of Belgium 
Avenue Circulaire, 3 
B-1180 Brussels BELGIUM}

\author[0000-0003-3571-8728]{Graham Barnes}
\affiliation{NorthWest Research Associates
3380 Mitchell Lane
Boulder, CO 80301 USA}

\author{Suzy Bingham}
\affiliation{Met Office
FitzRoy Road
Exeter, Devon, EX1 3PB, UNITED KINGDOM}

\author[0000-0002-4183-9895]{D. Shaun Bloomfield}
\affiliation{Northumbria University
Newcastle upon Tyne 
NE1 8ST, UNITED KINGDOM}

\author[0000-0002-4830-9352]{Aoife E. McCloskey}
\affiliation{School of Physics,
Trinity College Dublin,
College Green,
Dublin 2, IRELAND}

\author[0000-0001-5307-8045]{Veronique Delouille}
\affiliation{STCE - Royal Observatory of Belgium
Avenue Circulaire, 3
B-1180 Brussels BELGIUM}

\author{David Falconer}
\affiliation{NASA/NSSTC
Mail Code ST13
320 Sparkman Drive Huntsville, AL 35805, USA}

\author[0000-0001-9745-0400]{Peter T. Gallagher}
\affiliation{School of Cosmic Physics,
Dublin Institute for Advanced Studies,
31 Fitzwilliam Place,
Dublin, D02 XF86, IRELAND}

\author[0000-0001-6913-1330]{Manolis K. Georgoulis}
\affiliation{Department of Physics \& Astronomy
Georgia State University
1 Park Place, Rm \#715,
Atlanta, GA 30303, USA}
\affiliation{Academy of Athens
4 Soranou Efesiou Street, 11527 Athens, GREECE}

\author{Yuki Kubo}
\affiliation{National Institute of Information and Communications Technology\\
Space Environment Laboratory\\
4-2-1 Nukuikita Koganei Tokyo 184-8795 JAPAN}

\author[0000-0001-8969-9169]{Kangjin Lee}
\affiliation{Meteorological Satellite Ground Segment Development Center\\
Electronics and Telecommunications Research Institute, Daejeon \\
218 Gajeong-ro, Yuseong-gu, Daejeon, 34129, REPUBLIC OF KOREA}
\affiliation{Kyung Hee University \\
1732, Deogyeong-daero, Giheung-gu, Yongin, 17104, REPUBLIC OF KOREA}

\author{Sangwoo Lee}
\affiliation{SELab, Inc.
150-8, Nonhyeon-ro, Gangnam-gu, Seoul, 06049, REPUBLIC OF KOREA}

\author[0000-0001-5655-9928]{Vasily Lobzin}
\affiliation{Bureau of Meteorology
Space Weather Services
PO Box 1386 Haymarket NSW 1240 AUSTRALIA}

\author{JunChul Mun}
\affiliation{Korean Space Weather Center
198-6, Gwideok-ro, Hallim-eup, Jeju-si, 63025, REPUBLIC OF KOREA}

\author[0000-0002-9378-5315]{Sophie A. Murray}
\affiliation{School of Physics,
Trinity College Dublin,
College Green,
Dublin 2, IRELAND}
\affiliation{School of Cosmic Physics,
Dublin Institute for Advanced Studies,
31 Fitzwilliam Place,
Dublin, D02 XF86, IRELAND}

\author{Tarek A.M. Hamad Nageem}
\affiliation{University of Bradford
Bradford West Yorkshire BD7 1DP UNITED KINGDOM}

\author[0000-0002-8637-1130]{Rami Qahwaji}
\affiliation{University of Bradford
Bradford West Yorkshire BD7 1DP UNITED KINGDOM}

\author{Michael Sharpe}
\affiliation{Met Office
FitzRoy Road
Exeter, Devon, EX1 3PB, UNITED KINGDOM}

\author[0000-0001-8123-4244]{Rob Steenburgh}
\affiliation{NOAA/National Weather Service
National Centers for Environmental Prediction
Space Weather Prediction Center, W/NP9
325 Broadway, Boulder CO 80305 USA}

\author[0000-0002-9176-2697]{Graham Steward}
\affiliation{Bureau of Meteorology
Space Weather Services
PO Box 1386 Haymarket NSW 1240 AUSTRALIA}

\author[0000-0002-6290-158X]{Michael Terkildsen}
\affiliation{Bureau of Meteorology
Space Weather Services
PO Box 1386 Haymarket NSW 1240 AUSTRALIA}

%\begin{keypoints}
%\item Results are presented from a coordinated international comparison of operational solar flare forecasting systems; Paper II of a series.
%\item The behavior and performance of the methods are quantitatively evaluated in the context of broad implementation differences between the different methods.
%\item Including information on prior flaring improves the performance of the forecasts, as does having a ``forecaster in the loop'', with the strong caveat of relatively small sample sizes and weak trends; other trends are weaker but discussed.
%\end{keypoints}

\begin{abstract}
A workshop was recently held at Nagoya University (31 October -- 02
November 2017), sponsored by the Center for International Collaborative
Research, at the Institute for Space-Earth Environmental Research,
Nagoya University, Japan, to quantitatively compare the performance of
today's operational solar flare forecasting facilities.  Building upon
Paper I of this series \citep{allclear}, in Paper II \citep{ffc3_1}
we described the participating methods for this latest comparison effort,
the evaluation methodology, and presented quantitative comparisons.
In this paper we focus on the behavior and performance of the methods
when evaluated in the context of broad implementation differences.
Acknowledging the short testing interval available and the small
number of methods available, we do find that forecast performance: 1)
appears to improve by including persistence or prior flare activity,
region evolution, and a human ``forecaster in the loop''; 2) is hurt
by restricting data to disk-center observations; 3) may benefit from
long-term statistics, but mostly when then combined with modern data
sources and statistical approaches.  These trends are arguably weak
and must be viewed with numerous caveats, as discussed both here and
in Paper II.   Following this present work, we present in Paper IV
a novel analysis method to evaluate temporal patterns of forecasting
errors of both types \citep[{\it i.e.}, misses and false alarms; ][]{ffc3_3}.
Hence, most importantly, with this series of papers we demonstrate the
techniques for facilitating comparisons in the interest of establishing
performance-positive methodologies.

\end{abstract}

\keywords{methods: statistical -- Sun: flares -- Sun: magnetic fields}

\section{Introduction}
\label{sec:intro}

In 2009, the first in a series of workshops was held to compare and 
evaluate solar flare forecasting methods; the results and comparison
methodologies were presented in \citet[][; Paper I]{allclear} and have
informed numerous works.
In \citet[][`Paper II']{ffc3_1}, the initial results from the most recent 
``head-to-head'' comparison of operational flare-forecasting methods
are presented.  The comparison is the output of a  3-day workshop held
at the Institute for Space-Earth Environmental Research (ISEE) at 
Nagoya University over 31 October -- 02 November 2017, and was sponsored
by the ISEE  Center for International Collaborative Research.  In that
paper, the methodology was presented:  the agreed-upon testing interval,
event definitions, and evaluation metrics were described.  Specifically, 
daily operational full-disk forecasts from a variety of facilities
were gathered for 2016--2017, specifically for two event definitions:
\CC\ and \MM\ which indicate minimum-threshold for an event, 
the latency between forecast issuance and validity period start,
and the validity period itself.   The results
demonstrated broad performance similarities across numerous metrics for
the majority of methods.  The ``winner'' depended on event definition
and metric used.  However, within the estimated uncertainties, a more appropriate
description is that a number of methods consistently scored above
the ``no skill'' level. 

Simply comparing the performance is of limited use if there is no
investigation into ``why'', from which we may derive how improvements
could be made.   The question we investigate here is {\it ``are there
certain aspects, certain approaches or methodologies implemented by
the different methods that influence the performance in a discernible,
distinguishable way?''}

The participating facilities and methods (and their monikers and published
references, where available) are listed in Table~1 of Paper II, with
details that are not available by published literature are briefly
described in that paper's Appendix; an abbreviated version of that
table is reproduced here in Appendix~\ref{sec:method_table}.  
The submitted forecasts are also available\footnote{Leka and Park 2019, Harvard Dataverse, doi:10.7910/DVN/HYP74O}.
We take the descriptions further here, into the details of implementation that
the workshop group hypothesized may factor into performance.

\section{Methodology}
\label{sec:methods}

The approach here is to identify general categories by which the
methods could be grouped, and then examine whether there are systematic
performance differences according to those categories across a variety of
quantitative evaluation metrics.  As such, ``the devil is in the details''
and in most cases there was significant additional information needed
than what is readily available in the literature (see also Paper II,
Appendix A).

The participants wanted to determine whether implementation differences
could make a significant difference to the forecast performance.
In \citet{allclear}, this question was briefly investigated: we examined
the impact of subtle differences in how a commonly-used analysis quantity,
the total magnetic flux, was calculated ({\it e.g.}, any noise threshold
used, the specific de-projection method employed, if any).  could in fact
significantly impact the evaluation results.  For operational systems,
for example, one can imagine that restricting the relevant data analysis
to near-disk-center data will result in a systematic underperformance
in full-disk forecasts due to missing regions.  Were there any such
situations?  And what was the magnitude of such an impact?

Given the complexity of operational forecasting facilities, we asked
{\it ``at what other steps in the process were there multiple options
available?''}, and {\it ``is it possible to determine the impact of such
options on performance outcomes?''}  We identified four broad stages
at which differences arose: 1) the data used and how they are treated;
2) the specifics of training the method; 3) the specifics of producing the
forecast; 4) the actual issuance of the forecast itself.

All methods were requested to comment on specific questions regarding
particular aspects that were known to vary between methods which
the group felt may impact performance.  The topics and the responses
are summarized in Tables~\ref{tbl:questions1}--\ref{tbl:questions4}.
Some methods have multiple options for producing forecasts, and those
are delineated within the table.  Acronyms are used for brevity in the
tables and figures and some of the discussion, but are expanded upon 
in Appendix~\ref{sec:tlas}.

This approach will not capture all possible subtleties.  For example,
DAFFS and DAFFS-G may use a measure of prior flare activity with some event definitions but
not others, and this may change upon periodic re-training.
As another example, many methods use NOAA active-region designations,
others use HMI ``active region patches'' \citep[HARPs;][]{hmi_pipe}) that may or may not
agree in their entirety with the NOAA designations, while other methods
use various algorithms to independently determine solar magnetic regions.
Some of those methods have the goal of matching the NOAA designations,
but some algorithms perform region identification explicitly without that goal (such as the
HMI algorithm).  For the tests here, the region-assigned probabilities
for all regions were combined (generally by the methods themselves)
to produce full-disk probabilities, but questions linger as to how differences in region
determination impacts the training (upon which forecasts are then based).
Still, we attempt to answer what is answerable, or at least demonstrate
an appropriate methodology for doing so.

The metrics used here are the same as in Paper II \citep{ffc3_1}, representing a mix
of scores based on probabilistic and dichotomous forecasts.   For the latter, a single
probability threshold $P_{\rm th}=0.5$ is applied for the evaluation, and all other
considerations regarding the metric calculations discussed
in Paper II are applied here.  Essentially the 
individual scores have not changed from those presented in Paper II, but
what has changed is that each method is assigned membership to a particular
group (see Section~\ref{sec:groups}) and the resulting 
scores from within each group are presented together.  Instead of 
presenting the scores for each method individually, we emphasize
variation between categories by showing ``box \& whisker'' plots.

For the analysis here, two Paper II methods are generally excluded.  The first is the
120-day prior climatology forecast, an ``unskilled'' forecast that
can be constructed at the time of forecast issuance.  It was presented
in Paper II \citep[following][]{SharpeMurray2017} for evaluation across
the metrics, and used as the reference forecast for two skill
scores in order to specifically measure skill beyond a no-skill forecast
method.  In this analysis it is still used as a reference forecast
for the ApSS\_clim and MSESS\_clim metrics, however it is not presented on
its own for evaluation (as was done in Paper II), because we focus here
on methods that hopefully bring added value beyond an unskilled method.

The second method excluded from the quantitative analysis is the
NJIT method.  As discussed in Paper II, the NJIT method represents
a research project that was never fully transitioned to operations,
and as such suffers in numerous metrics from missing forecasts; it 
is a consistent outlier.  Again, with a focus on operational methods,
for this analysis we omit the NJIT forecasts when computing the metrics
(although we include its details in the Tables for future reference).

%\clearpage
\begin{longrotatetable}
\begin{deluxetable}{p{4cm}p{16cm}}
\tabletypesize{\scriptsize}
\tablewidth{19cm}
\tablecaption{Devil-is-in-the-Details Summary\label{tbl:questions1}\\
{\bf Forecast Data Sources and Treatment:} What are primary, backup data sources?  Is there a protocol for bad / missing data?  If using $B_{\rm los}$, are any corrections used?  Are there limits on the data?  Is there any special treatment of the data?} 
\tablehead{\colhead{Method} & \colhead{Response} }
\startdata
%\multicolumn{2}{l}{{\bf Forecast Data Sources and Treatment:} What are primary, backup data sources?  Is there a protocol for bad / missing data?  If using $B_{\rm los}$, are any corrections used?  Are there limits on the data?  Is there any special treatment of the data?} \\ 
%
A-EFFORT\dotfill & HMI NRT FD $B_{\rm los}$ data, $B_r^{los} = B_{\rm los} / \cos(\theta)$ \& heliographic-plane 
projection, HMI-to-MDI emulation, NOAA SRS AR assignments; missing data protocol: prior forecast does not refresh\\
AMOS\dotfill & NOAA-reported flare events and NOAA SRS AR reports (2 days' worth); missing data protocol: prior forecast does not refresh. \\
ASAP\dotfill & HMI NRT FD $B_{\rm los}$ \& Continuum;  No protocol for missing data; not using $B_{\rm los}$ 
quantitatively (region identification only).  \\ 
ASSA\dotfill & HMI NRT FD $B_{\rm los}$ \& Continuum; No protocol for missing data.   No correction to $B_{\rm los}$ but 
sunspots located $>80^\circ$ from the limb are excluded. \\
BOM\dotfill & NOAA/SWPC SRS, USAF SOON reports, HMI NRT $B_{\rm los}$ rebinned by x4; replaced by definitive data 
after a few days (for future training);  bad/missing data protocol: reverts to forecasts by
region classification / area / flare rates. \\
DAFFS, DAFFS-G\dotfill & HMI NRT $\vec{\bf B}$ and NRT HARP designations, NOAA NRT GOES-based X-ray event lists (DAFFS), 
GONG $B_{\rm los}$ + GOES for DAFFS-G, used when HMI data not available; if neither
HMI or GONG are available, GOES X-ray events used with NOAA AR designations; training-interval 
climatology as last resort.  $B_{\rm los}$ data: uses $B_r^{\rm pot}$ estimate \citep{bbpot}. \\ 
MAG4\dotfill & HMI NRT FD $B_{\rm los}$ (GONG manually as backup; not employed here), $\vec{\bf B}$ data, 
NOAA SRS AR assignments, NOAA-reported flare events; LMSAL/SolarSoft events as back-up.  
Use last good data up to 60-96 minutes delay, else repeat last forecast.  Prior flaring (MAG4*F) set
to null if data are unavailable.  No correction to $B_{\rm los}$.  Limits imposed on training data (see 
Table~\ref{tbl:questions3}). \\
MCSTAT, MCEVOL\dotfill & NOAA flare event and SRS reports (Zpc classes); missing SRS report protocol: 0\% forecast. \\
MOSWOC\dotfill & HMI images, SWPC AR numbers.  {\it SDO} data used qualitatively.  \\
NICT\dotfill & NOAA SRS \& GOES, HMI imaging data, HMI SHARP parameters, AIA imaging data, ground-based data 
as backup.  {\it SDO} data used qualitatively.\\
NJIT\dotfill & NOAA SRS \& HMI $B_{\rm los}, \cos(\theta)$ correction; helicity is not computed (and no 
forecast is issued) if NRT data are not downloaded or available in the NJIT flare forecasting system 
(for any reason). \\
NOAA\dotfill & NOAA \& USAF imagery, flare reports, radio data; any \& all imagery, primarily NOAA-assured operational sources (including GOES, GONG assets), other as needed/available.  {\it SDO} data used qualitatively.  No protocol for outtages beyond ``any and all'' data used.\\ 
SIDC\dotfill & NOAA SRS and Catania Obs; GOES flare history (PROBA2/LYRA as backup); {\it SDO}/HMI magnetogram and continuum movies, EUV images ({\it SDO}/AIA, {\it PROBA2}/SWAP as backup, and {\it STEREO}/EUVI), especially for limb-ward regions.  \\ \hline
\enddata
\end{deluxetable}
\end{longrotatetable}

\begin{longrotatetable}
\begin{deluxetable}{p{4cm}p{16cm}}
\tabletypesize{\scriptsize}
\tablewidth{19cm}
\tablecaption{Devil-is-in-the-Details Summary (cont'd)\label{tbl:questions2} \\
{\bf Full-Disk Forecast Production:} How are active regions identified? How are full-disk forecasts constructed?
Is there any explicit forecasting for behind-the-limb events?} 
%
%\multicolumn{2}{p{20cm}}{{\bf Full-Disk Forecast Production:} How are active regions identified? How are full-disk forecasts constructed?  
%Is there any explicit forecasting for behind-the-limb events?} \\ \hline
\tablehead{\colhead{Method} & \colhead{Response} }
\startdata
A-EFFORT\dotfill & Regions IDd via ARIA \citep{LaBonte_etal_2007,Georgoulis_etal_2008}; FD forecasts via region probabilities. No behind-limb forecasts\\
AMOS\dotfill &  Regions ID'd by NOAA/SRS files; FD forecasts via region probabilities.  \\
ASAP\dotfill & ML code to identify / classify sunspot regions using intensity and $B_{\rm los}$ images.  No full-disk 
prediction (region only).  \\
ASSA\dotfill & In-house automatic ID and classification of McIntosh \& Mt Wilson Classes.
%, plus spot grouping, asymmetry, and growth (12hr forecasts). 
Probabilities from classification, Poisson statistics.  FD forecasts via region probabilities.  \\
BOM\dotfill & Automatic recognition of ARs by magnetogram flux thresholds, NOAA/SRS and USAF/SOON as backup. 
FD forecasts via region probabilities. 
No explicit behind-the-limb forecasts or multi-day forecasts, 
although very-near limb regions assigned region-flaring climatology.\\
DAFFS, DAFFS-G\dotfill & HARPs (HMI, for DAFFS) or NOAA NRT region-based areas (GARPS, for GONG for DAFFS-G)
ID'd, extracted.  FD forecasts via region probabilities.  No explicit behind-the-limb forecasts 
beyond multi-day forecasts. \\ 
MAG4\dotfill & NOAA/SRS ARs, FD forecasts via region probabilities.  No explicit behind-the-limb forecasts beyond multi-day forecasts.\\
MCSTAT, MCEVOL\dotfill & NOAA/SRS ARs reports \\
MOSWOC\dotfill & NOAA/SRS ARs plus additional regions ID'd and assigned Mt Wilson \& McIntosh classes if needed,
updated 4x/day.  FD forecasts via region probabilities. No explicit behind-the-limb forecasts beyond multi-day forecasts\\
NICT\dotfill & NOAA/SRS information is used internally, but FD forecasts only are issued. \\
NJIT\dotfill & NOAA/SRS used for AR identification, FD forecasts via region probabilities. \\
NOAA\dotfill & NOAA/SWPC produces region identification and classification, disseminates.  FD forecasts via region probabilities.  Forecasts include probabilities for behind-the-limb activity.\\
SIDC\dotfill & Catania Region identification \& NOAA/SRS for region probabilities then FD forecasts via region probabilities, human modified ({\it e.g}, for new or behind-the-limb regions) \\ \hline
\enddata
\end{deluxetable}
\end{longrotatetable}
\begin{longrotatetable}
\begin{deluxetable}{p{4cm}p{16cm}}
\tabletypesize{\scriptsize}
\tablewidth{19cm}
\tablecaption{Devil-is-in-the-Details Summary (cont'd)\label{tbl:questions3} \\
{\bf Training:} What data are used, what is optimized / produced?  Are balanced
training sets imposed or is class (event / no-event) imbalance accomodated?  What interval is used in general / for this test (if different)?
Is there a protocol for training for behind-the-limb or unassigned events?} 
%
%\multicolumn{2}{p{20cm}}{{\bf Training:} What data are used, what is optimized / produced?  Are balanced 
%training sets imposed?  What interval is used in general / for this test (if different)?  
%Is there a protocol for training for behind-the-limb or unassigned events?} \\ \hline
\tablehead{\colhead{Method} & \colhead{Response} }
\startdata
A-EFFORT\dotfill & Forecasts curves constructed, no further optimization.  80\% of calendar days of archive data, contiguous or random select.  3-hr forecast cadence for first 12 months of servce; balancing in training to a 4:1 (time-span), climatological sample ratios.   \\
AMOS\dotfill & 1996 - 2010 McIntosh class flaring rate, probabilities from historical McIntosh rates plus 
factor for sunspot area change in prior 24hr via Poisson statistics. \\
ASAP\dotfill & Trained on ASAP-produced sunspot ID's and associated flare events 1982 -- 2013; Neural nets optimized 
on mean square error (MSE). \\
ASSA\dotfill & Training on MDI and HMI data, generally MDI and HMI data 1996 -- 2013 (Zpc-forecasts).
A change in training occurred during the testing interval: 2016.01.01 -- 2016.12.18 were trained
with 1996--2010 {\it SOHO}/MDI data, and then 2016.12.19 -- 2017.12.31 were trained using 1996--2010 SOHO MDI 
and 2011--2013 {\it SDO}/HMI data. \\
BOM\dotfill & Automated Active Region detection optimized to match SRS reports 2011--2015; 
Flarecast II (logistic regression model): HMI definitive $B_{\rm los}$ 2010.05.01 -- 2015.12.31 used
for training, variables selected to minimize Aikake's Information Criteria (AIC) 
and LRM uses maximum likelihood to estimate the coefficients of the model.  All HMI definitive data used
2010.05.01 -- 2015.12.31, naturally unbalanced.  No training for behind-the-limb.  \\
DAFFS, DAFFS-G\dotfill & Training from HMI NRT era until designated date (2012.10.22 -- 2015.12.31 for this workshop), or 
GONG era (2006.09.01 -- 2015.12.31); X-Ray events for prior flare activity parameters trained with the magnetic 
source data (matching that training interval).  Parameter pair(s) can change
upon re-training and will vary between event definitions.
Events not identified with regions are ignored.  
DAFFS* trains to optimize Brier Skill Score. \\ 
MAG4\dotfill & MDI interval (1996 -- 2004), plus HMI-to-MDI degradation of HMI data.  Training
data limits relative to CM: 30$^\circ$ ($B_{\rm los}$); 60$^\circ$ ($\vec{B}$).  Probabilities
derived from event rates after fitting free-energy proxy to empirical event rate curves.\\
MCSTAT, MCEVOL\dotfill & Both: No behind-the-limb events considered.  No correction for class imbalance.
Poisson statistics produce probabilistic forecasts.
MCSTAT: 1969-1976 [M- \& X-class] (SC 20) plus Dec 1988 -- Jun 1996 [C, M \& X-class] (SC 22).  
MCEVOL: 1988-1996 (SC 22) plus 1996-2008 (SC 23).  
Poisson rates trained from 24\,hr changes between full McIntosh classifications by counting
\# flares within 24\,hr following a classification change.  Evolution computed within $\pm75^\circ$
of CM to avoid limb-affected misidentification in training. \\
MOSWOC\dotfill & Initial forecast probabilities based on historical rates and McIntosh classes 1969--2011. \\
NICT\dotfill & Human training on self-validation results from 1992 onward.\\
NJIT\dotfill & 1996/01/01 -- 2006/12/31; No Behind-the-limb events used for training, no consideration for class
imbalance.  Probabilities are based on forecast curves from training data.\\
NOAA\dotfill & Initial forecast probabilities based on historical rates and McIntosh classes 1969--2011. \\ 
SIDC\dotfill & Probabilities from historical rates and McIntosh classes (SC 22 1988--1996) assuming Poisson statistics. \\ \hline
\enddata
\end{deluxetable}
\end{longrotatetable}

\begin{longrotatetable}
\begin{deluxetable}{p{4cm}p{16cm}}
\tabletypesize{\scriptsize}
\tablewidth{19cm}
\tablecaption{Devil-is-in-the-Details Summary (cont'd) \label{tbl:questions4} \\
{\bf Forecasts:} Are humans involved and if so, how?
How are forecasts produced from the data?
Is there a behind-the-limb protocol for forecasts?
Is there a single forecast or additional customised forecasts?
Are there restrictions (distance from disk center, size of region, data quality, {\it etc.}),
and if so, what is used in its place (e.g., climatology)?}
%
%\multicolumn{2}{p{20cm}}{{\bf Forecasts:} Are humans involved and if so, how?  
%How are forecasts produced from the data?  
%Is there a behind-the-limb protocol for forecasts?
%Is there a single forecast or additional customised forecasts? 
%Are there restrictions (distance from disk center, size of region, data quality, {\it etc.}),  
%and if so, what is used in its place (e.g., climatology)?} \\ \hline
\tablehead{\colhead{Method} & \colhead{Response} }
\startdata
A-EFFORT\dotfill & AR ID and $B_{\rm eff}$ calculation is limited to $50^\circ$; ARs located $50^\circ - 70^\circ$ from CM: a proxy is used: $B_{\rm eff} = 10^{-21.961396}\Phi_{\rm tot}^{1.0834181}$.  
Processing $> 45^\circ$ from CM problematic.  No behind-limb forecasts, 
24-hr validity, 3-hr refresh, 0-hr latency, for \Mp, \Mfp, \Xp, \Xfp; email alerts issued upon request.\\
AMOS\dotfill & No behind-limb forecasts, no humans in the loop,  \Co, \Mo (not exceedance), and \Xp\ for
each NOAA AR and full-disk, 24h validity, 0hr latency.\\
ASAP\dotfill & Region forecasts for 6, 12, 24, 48 hour validity periods, \Mo (not exceedance), and \Xp.\\
ASSA\dotfill & Hourly refresh, no human, for 24hr validity (Zpc-based).
Forecasts issued for \Co, \Mo (not exceedance), and \Xp;  no behind-limb forecasts.  \\
BOM\dotfill & A logistic regression model (LRM) is used to generate \Mp, \Xp, region and full-disk, 
probabilistic \& deterministic forecasts (per customer specifications) for flaring activity over the next 
24\,hr updated at 00, 06, 12, 18 UT. \\
DAFFS, DAFFS-G\dotfill & No humans, behind-limb forecast indirectly through longer-range forecasts.   
Magnetogram data limit: $\pm 84^\circ$. 
Discriminant analysis (training) provides best-performing parameter pairs and their PDEs which forecast probabilites derived.
Forecasts: 24h validity, 0, 24, 48hr latencies, \Cp, \Mp, \Xp issued @11:54 and 23:54 UT. 
Customized cost-based forecasts and forecasts for different event definitions available.  \\
MAG4\dotfill & Warnings issued for forecasts using data beyond training limits; no behind-limb forecasts.
\Mp, \Xp, 24h validity, 0h latency (effectively).  Four modes (``MAG4W'',``MAG4WF'',``MAG4VW'',``MAG4VWF'')
according to permutations of $B_{\rm los}$, ``de-projected'' $\vec{B}$, and previous flare history.
Regions with area beyond $85^\circ$ are not included;
forecasts are provided to $\pm 85^\circ$ but with warnings beyond $45^\circ$ that event rate probabilities 
may be underestimated.  All four forecasts available through {\tt https://www.uah.edu/cspar/research/mag4-page}. \\
MCSTAT, MCEVOL\dotfill &  MCSTAT: No limit (full visible disk).  MCEVOL: No limit (full visible disk). \\
MOSWOC\dotfill & Human forecaster modifies probability from Poisson statistics, including considerations
for flaring history and indications of flare potential from not-visible regions.  24-hr forecasts for 
0, 24, 48, 72-hr latencies for \Mo (not exceedance), \Xp\ at 00:00 and 12:00UT (latter is a 12-hr ``updated' forecast). \\
NICT\dotfill & 4-category 24hr deterministic forecasts (max class of {\tt A1.0--B9.}, of \Co, of \Mo, or of \Xp), 
at 06:00 daily; human-based forecast. \\
NJIT\dotfill & Regions included within $\pm 60^\circ$ from CM.  Forecast for \Co, \Mo
(not exceedance), \Xp\ maximum class.\\
NOAA\dotfill & Human forecaster modifies probability from region-class climatology.  Behind-the-limb events 
included in forecast based on AR-based flare persistance.  Exceedance forecasts of \Cp, \Mp, \Xp, 24h 
validity for 0, 24, 48hr latency, issued at 22:00 with possible updates to 00:30-issued ``3-day forecast 
product'' (with further updates as needed for second issuance at 12:30).  Those forecasts not publicly archived 
as the ``RSGA'' data product are available internally ({\it e.g.} \Cp\ forecasts).\\ 
SIDC\dotfill & Human forecaster modifies probability from Poisson statistics. Issue time: 12:30UT, 24hr validity.
Exceedance probabilities for \Cp, \Mp, \Xp\ flares, per active regions and FD. 
Away from CM, data sources other than $B_{\rm los}$ are used.  \\ \hline
\enddata
\end{deluxetable}
\end{longrotatetable}

\subsection{Broad Characteristics Groupings}
\label{sec:groups}

The goal of this analysis is to identify broad characteristics of the
forecasting methods that provide improved performance.   We identified
a few tenable categories for analysis, described below.  Some of
the characterizations are straightforward (such as whether, and in
what manner, persistence or prior flaring activity is included), while others are more subtle
or may not exactly describe the differences between implementation.
In that manner, assignments were made by the method representatives
(see Table~\ref{tbl:categories}) and any caveats to that assignment
should be covered in Tables~\ref{tbl:questions1}--\ref{tbl:questions4}.
The results for each grouping are presented in an associated figure, and 
discussed further in section~\ref{sec:results}.

\noindent
\paragraph*{Training Interval} (Figure~\ref{fig:comp_train}):  The
difference in the length of the training interval was specifically
targeted for this categorization.  Generally speaking, the methods
relying solely on ``high quality data'' such as from the Solar Dynamics
Observatory Helioseismic and Magnetic Imager 
\citep[{\it SDO}/HMI;][, see acronyms in Appendix~\ref{sec:tlas}]{sdo,hmi_cal,hmi_invert,hmi_pipe} were
considered to have employed ``Short'' training intervals, compared to
those using longer baselines of information (such as more than one solar
cycle's worth of McIntosh classifications and the associated flaring rates
\citep{McIntosh1990}) that were assigned as ``Long''.  Additionally there
were ``Hybrid'' systems.  These may use modern data for the forecasts but
were trained on other data so as to take advantage of a longer baseline,
with some calibration performed between the two.  Alternatively,
members of this ``hybrid'' category merged forecasts from multiple
systems with different training intervals available.  The ``Short''
category was the minority. 
%The results in Figure~\ref{fig:comp_train}
%summarize the performance according to the assignments that are indicated
%in Table~\ref{tbl:categories}.

\noindent
\paragraph*{Forecast Production} (Figure~\ref{fig:comp_fcastmethod}):
This classification refers specifically to the statistical method employed
in order to relate the training period and training data to the new data
and the method used to produce the actual forecast from said new data.
We identified three sub-categories.  First,  ``Machine Learning /Classifier''
employs a statistical classifying approach to the training analysis
and to producing the forecast.  Second, ``Not Machine Learning'' uses empirical
fitting to historical data including approaches such as regression
curves, Poisson-statistics analysis of flaring rates according to sunspot
region classification schemes, further conversion from flaring rates to
probabilities, {\it etc.}.  Finally, for the ``forecaster in the loop''
(FITL) designation results may be obtained with or without
either of the other two approaches, but are then routinely adjusted
or assimilated with other human input to produce a final forecast.
%The results in Figure~\ref{fig:comp_fcastmethod} summarize the performance
%according to the assignments from Table~\ref{tbl:categories} with the
%acknowledgement that FITL may or may not implicitly include influence
%from the other categories.

\noindent
\paragraph*{Observational Limits / Forecast Extent} (Figure~\ref{fig:comp_zones}): 
This categorization pertains to the data used when calculating the
forecasts (without explicit reference to the training).  Some methods
limit the data used for the forecasts to only those which lie close to
the central meridian (CM); we call these ``Restricted'' if the limit
is stricter than essentially on or nearly approaching the limb ({\it
i.e.} $< \approx 80^\circ$ from disk center).  Other methods effectively use
data from the full visible disk without significant restriction and we
call these ``Visible Disk''; this is by far the most popular category.
Both of these categories only forecast flares from visible regions (except
in cases of longer-range forecasts for limb-approaching regions, which
are not considered here).  Finally, some methods include information on
not-yet-visible but expected regions (new or returning), or explicitly
project or extrapolate information for newly rotated-off regions for
``Earth Impacting'' forecasts -- in other words, forecasting for anything
impactful even from regions that are not yet or no longer visible.

\noindent
\paragraph*{Data Characterization} (Figure~\ref{fig:comp_params}):
The methods were first  divided into two broad groups, those
employing ``Simple'' parameters {\it vs.} those using ``Magnetic /
Modern Quantification''.  The former are generally McIntosh or Hale
classifications (or similar qualitative indicies) and are by-and-large
discrete assignments.  The latter are generally quantitative measures
generated from input quantitative data (primarily magnetic field data),
and are by-and-large continuous variables.  The first group included some
refinements between those that use the NOAA-(or other source)-determined
assignments and those which determined the classifications by their own
methods (including machine-learning based algorithms).  Those refinements
are indicated in the table notes, but are not included in
the further analysis shown in Figure~\ref{fig:comp_params}.

\noindent
\paragraph*{Persistence or Prior Flare Activity}
(Figure~\ref{fig:comp_persist}):  One significant difference between
methods is whether or not prior flaring activity is explicitly included;
many methods do not include it.  The term ``persistence'' specifically
means forecasting the same conditions as the present, and is somewhat
distinct from accounting for and including a measure of prior flare
activity over a specified interval.  Of those that do include one of
these measures, we distinguish in Table~\ref{tbl:categories} between
``automated'' algorithms (which, for example, quantitatively parametrize
prior flaring rates and include it in training as well as forecasting)
and those methods that use ``other'' ways to include the information,
such as the training of human forecasters (in which case the influence
of persistence information on the forecasts is generally qualitative).
In further analysis (see Figure~\ref{fig:comp_persist}) these refinements
are combined (and referred to simply as persistence) in order to show a `yes/no' comparison.

\noindent
\paragraph*{Evolution}  (Figure~\ref{fig:comp_evol}):  The evolution
of sunspot groups, in particular the rapidity of their growth or
decay, has long been recognized as a signal of higher flaring activity
\citep[{\it e.g.},][]{flareprediction,Lee_etal_2012,McCloskey_etal_2016}.
We distinguish between three approaches here: 1) no inclusion of evolution;
2) a quantitative analysis of evolution that is invoked during training
as well as for the forecast; 3) a qualitative inclusion of evolution
(most common for the FITL methods).  The methods are categorized
thus in Table~\ref{tbl:categories}, but in the accompanying
Figure~\ref{fig:comp_evol} these reduce to a `yes/no' assignment.

\begin{longrotatetable}
\begin{deluxetable}{l|ccc|ccc|ccc|cc|ccc|ccc|}
\tablecaption{Broad Characteristics\label{tbl:categories}}
%\begin{center}
%\begin{tabular}{l|ccc|ccc|ccc|cc|ccc|ccc|}
%
\tablehead{\multicolumn{18}{c}{}}
\startdata
& \multicolumn{3}{c|}{Training Interval} & \multicolumn{3}{c|}{Forecast Production} & 
\multicolumn{3}{c|}{Limits and Extent} & \multicolumn{2}{c|}{Data Characteristics} & 
\multicolumn{3}{c|}{Persistence} & \multicolumn{3}{c|}{Evolution} \\ 
Method  & \rotatebox[origin=r]{90}{Long} &  \rotatebox[origin=r]{90}{Short} &  \rotatebox[origin=r]{90}{Hybrid} & 
  \rotatebox[origin=r]{90}{ML/Classifier} & \rotatebox[origin=r]{90}{Not ML} & \rotatebox[origin=r]{90}{FITL} &
  \rotatebox[origin=r]{90}{Earth-Impacting} & \rotatebox[origin=r]{90}{FullDisk} & \rotatebox[origin=r]{90}{Restricted} &
 \rotatebox[origin=r]{90}{Simple} & \rotatebox[origin=r]{90}{Magnetic/Modern} &
 \rotatebox[origin=r]{90}{None} & \rotatebox[origin=r]{90}{Auto} & \rotatebox[origin=r]{90}{Other} &
 \rotatebox[origin=r]{90}{None} & \rotatebox[origin=r]{90}{Quantitative} & \rotatebox[origin=r]{90}{Qualitative} \\ \hline
A-EFFORT & & & $\bullet$ & & $\bullet$ & & & &$\bullet$ & & $\bullet$ & $\bullet$ & & & $\bullet$ & &\\
AMOS & $\bullet$ & & & & $\bullet$ & & & $\bullet$ & & $\bullet$ & & $\bullet$ & & & & $\bullet$ & \\
ASAP & & & $\bullet$ & $\bullet$ & & & & & $\bullet$ & $\bullet^{\dagger\ast}$ & & $\bullet$ & & & $\bullet$ & & \\
ASSA & & & $\bullet$ & & $\bullet$ & & & & $\bullet$ & $\bullet^{\dagger\ast}$ & & $\bullet$ & & & $\bullet$ & & \\
BOM & & $\bullet$ & & $\bullet$ & & & & $\bullet$ & & & $\bullet$ & & $\bullet$ & & $\bullet$ & & \\
DAFFS & & $\bullet$ & & $\bullet$ & & & & $\bullet$ & & & $\bullet$ & & $\bullet$ & & $\bullet$ & & \\
DAFFS-G & $\bullet$ & & & $\bullet$ & & & & $\bullet$ & & & $\bullet$ & $\bullet$ & $\circ$ & & $\bullet$ & & \\
MAG4W & & & $\bullet$ & & $\bullet$ & & & $\bullet^{\Uparrow}$ & & & $\bullet$ & $\bullet$ & & & $\bullet$ & & \\
MAG4WF & & & $\bullet$ & & $\bullet$ & & & $\bullet^{\Uparrow}$ & & & $\bullet$ & & $\bullet$ & & $\bullet$ & & \\
MAG4VW & & & $\bullet$ & & $\bullet$ & & & $\bullet^{\S}$ & & & $\bullet$ & $\bullet$  & & & $\bullet$ & & \\
MAG4VWF & & & $\bullet$ & & $\bullet$ & & & $\bullet^{\S}$ & & & $\bullet$ & & $\bullet$ & & $\bullet$ & & \\
MCEVOL & $\bullet$ & & & & $\bullet$ & & & $\bullet$ & & $\bullet$ & & $\bullet$ & & & & $\bullet$ & \\
MCSTAT & $\bullet$ & & & & $\bullet$ & & & $\bullet$ & & $\bullet$ & & $\bullet$ & & & $\bullet$ & & \\
MOSWOC & $\bullet$ & & & & & $\bullet$ & $\bullet$ & & & $\bullet^{\ast}$ & & & & $\bullet$ & & & $\bullet$ \\
NICT & $\bullet$ & & & & & $\bullet$ & $\bullet$ & & & $\bullet$ & & & & $\bullet$ & & & $\bullet$ \\
NJIT & & & $\bullet$ & & $\bullet$ & & & & $\bullet$ & & $\bullet$ & $\bullet$ & & & $\bullet$ & & \\
NOAA & $\bullet$ & & & & & $\bullet$ & $\bullet$ & & & $\bullet$ & & & & $\bullet$ & & & $\bullet$ \\
SIDC & $\bullet$ & & & & & $\bullet$ & $\bullet$ & & & $\bullet$ & & & & $\bullet$ & & & $\bullet$ \\ \hline\hline
\enddata
\begin{flushleft}
$\bullet$: Present / represented in submitted forecasts \\
$\circ$: Capability present but not invoked in all event definitions \\
$\ast$: Determines own reckoning of McIntosh class \\
$\dagger$: Determined by machine learning \\
$\Uparrow $: Forecasts issued with warnings for regions beyond $30^\circ$ \\
$\S$: Forecasts issued with warnings for regions beyond $45^\circ$
\end{flushleft}
\end{deluxetable}
\end{longrotatetable}
\normalsize
\clearpage

\section{Results}
\label{sec:results}

Citing performance metrics is becoming standard practice for published
research on event forecasting.  Herein we present the same evaluation
metrics described and calculated in Paper II, but with discrimination
according to the categories described above in an attempt to establish
the causes behind performance differences.

\begin{figure}[h]
\includegraphics[width=0.9\textwidth,clip, trim = 0mm 0mm 0mm 0mm, angle=0]{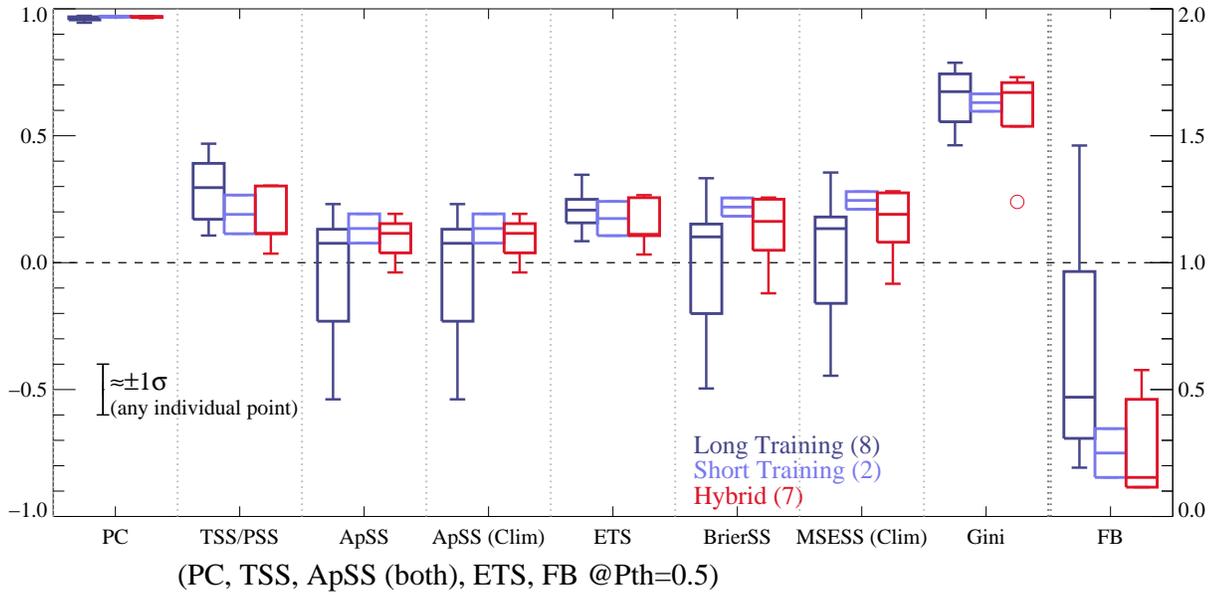}
\includegraphics[width=0.9\textwidth,clip, trim = 0mm 0mm 0mm 0mm, angle=0]{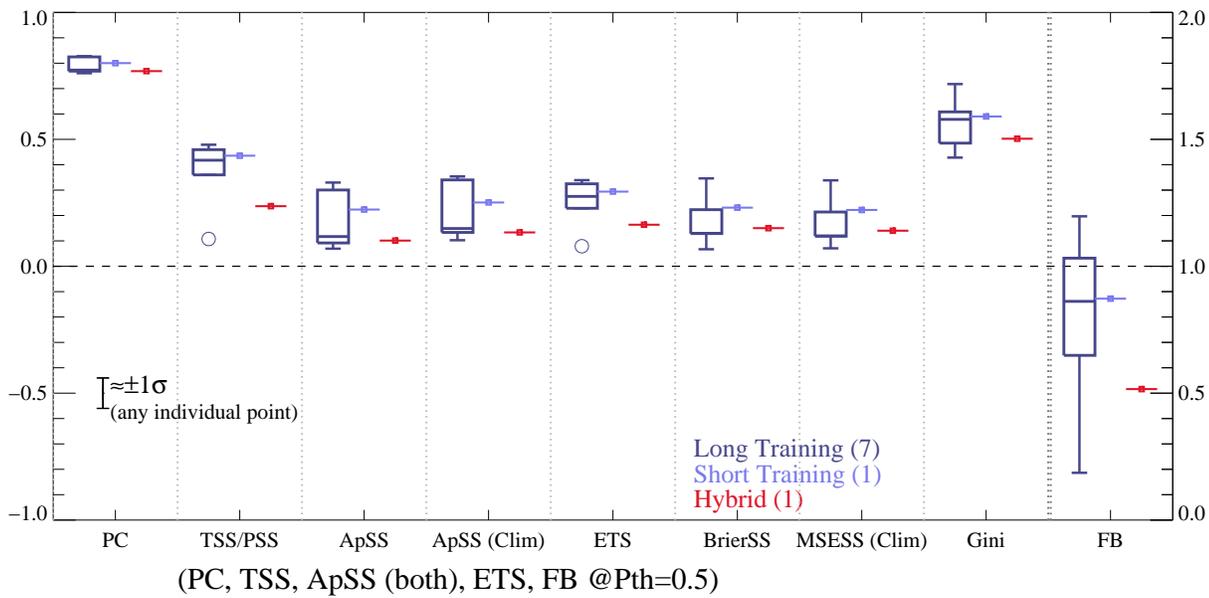}
\caption{Results from the direct comparison of flare forecasting methods,
as grouped by differences in the training interval used, as indicated,
for the \MM\ event definition (top) and \CC\ event definition (bottom).
Box \& whisker plots are used here, with the mid-line indicating the
median, boxes indicating the 25th and 75th percentiles, the whiskers
indicating maximum and minimum except circles showing those points
beyond 1.5$\times$IQR (the InterQuartile Range).  The number of methods
represented in each category is indicated with the category color/label.
The metrics are those described and presented in Paper II; of note, the Frequency
Bias (FB) is on a different scale, referencing the axis on the right.  CLIM120 and
NJIT are not included in this graphical analysis (see text).  Fewer
methods provide \CC\ forecasts, hence the sparseness of the points
relative to the \MM\ event definition. }
\label{fig:comp_train}
\end{figure}

\begin{figure}[h]
\includegraphics[width=0.9\textwidth,clip, trim = 0mm 0mm 0mm 0mm, angle=0]{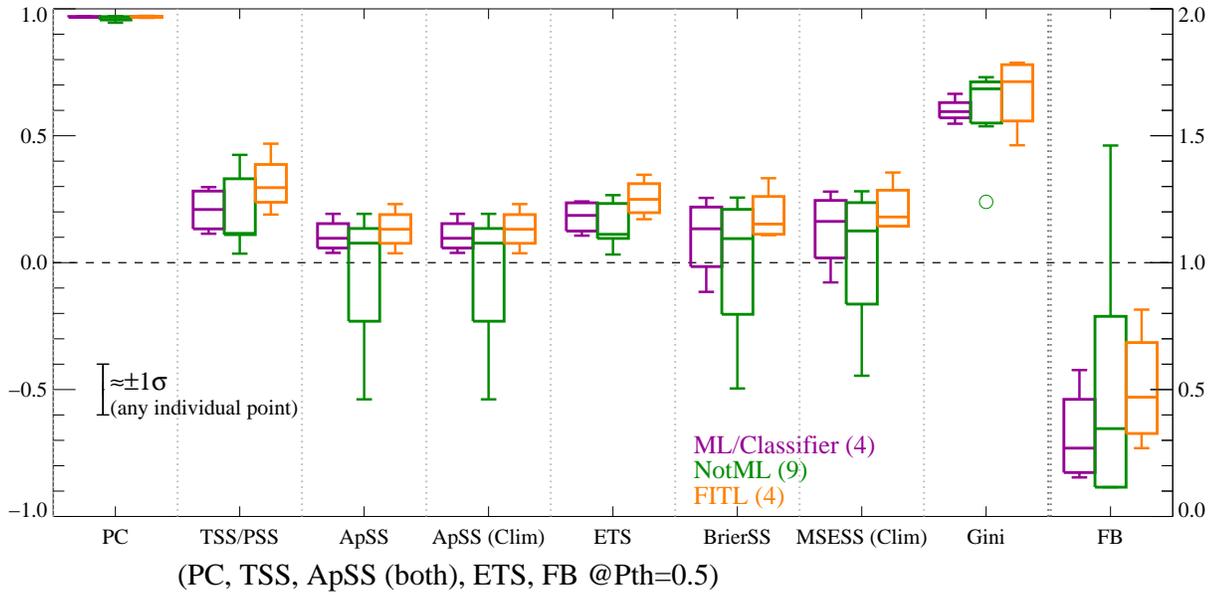}
\includegraphics[width=0.9\textwidth,clip, trim = 0mm 0mm 0mm 0mm, angle=0]{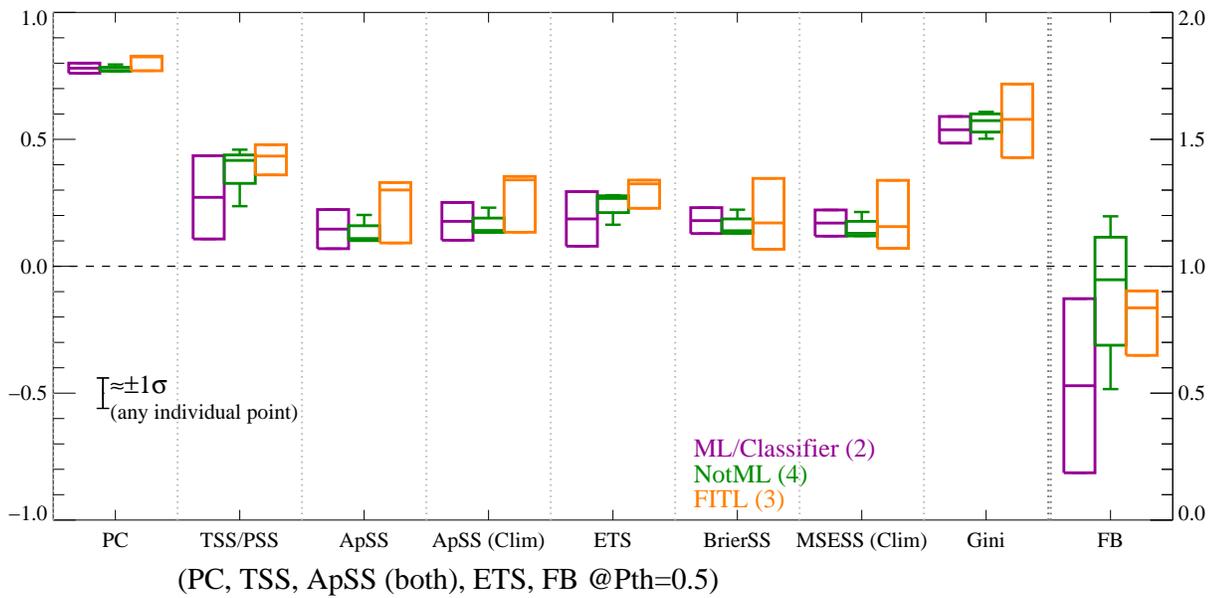}
\caption{Same as Figure~\ref{fig:comp_train}, but for comparisons of the methods by 
which the forecasts are produced, as indicated.}
\label{fig:comp_fcastmethod}
\end{figure}

\begin{figure}[h]
\includegraphics[width=0.9\textwidth,clip, trim = 0mm 0mm 0mm 0mm, angle=0]{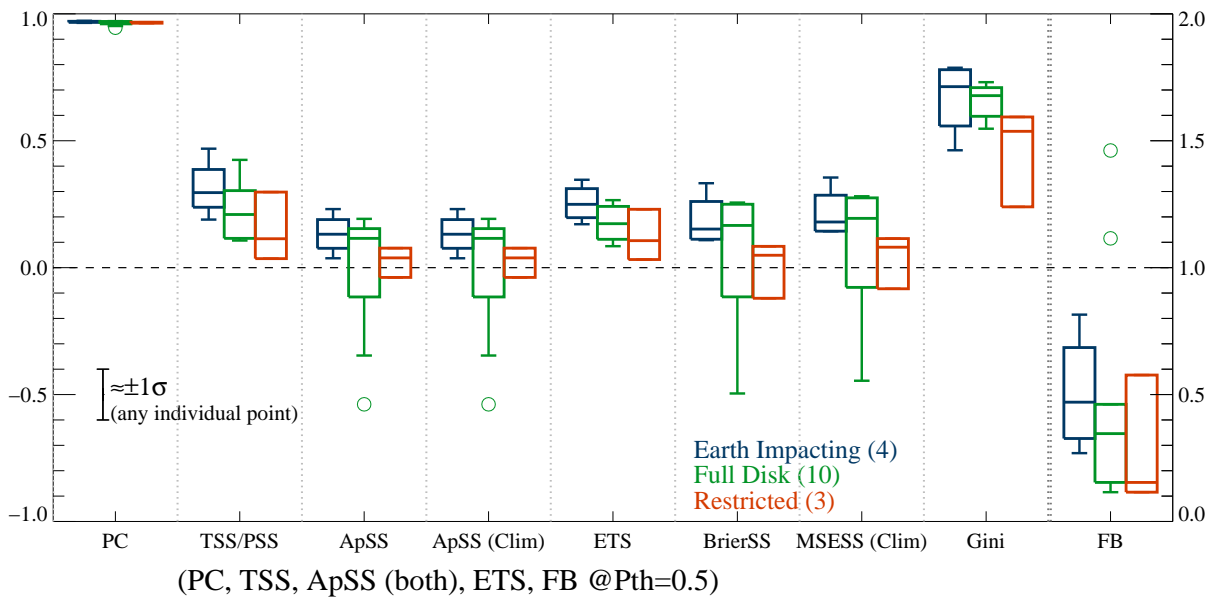}
\includegraphics[width=0.9\textwidth,clip, trim = 0mm 0mm 0mm 0mm, angle=0]{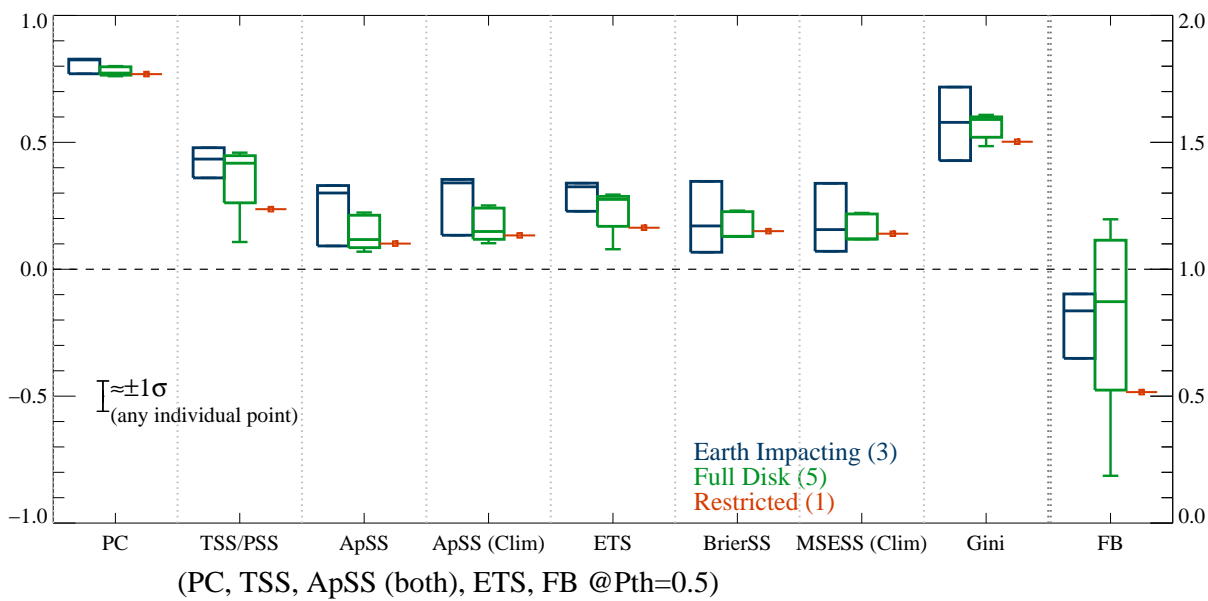}
\caption{Same as Figure~\ref{fig:comp_train}, but for comparisons of the zones for which a forecast is issued,
as indicated.}
\label{fig:comp_zones}
\end{figure}

\begin{figure}[h]
\includegraphics[width=0.9\textwidth,clip, trim = 0mm 0mm 0mm 0mm, angle=0]{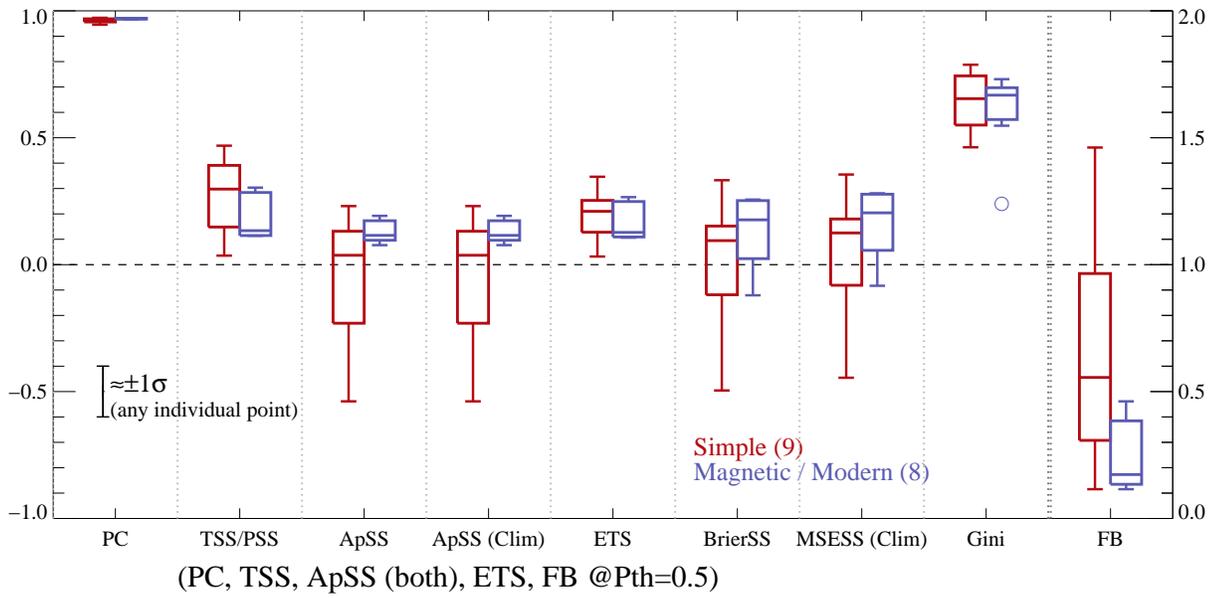}
\includegraphics[width=0.9\textwidth,clip, trim = 0mm 0mm 0mm 0mm, angle=0]{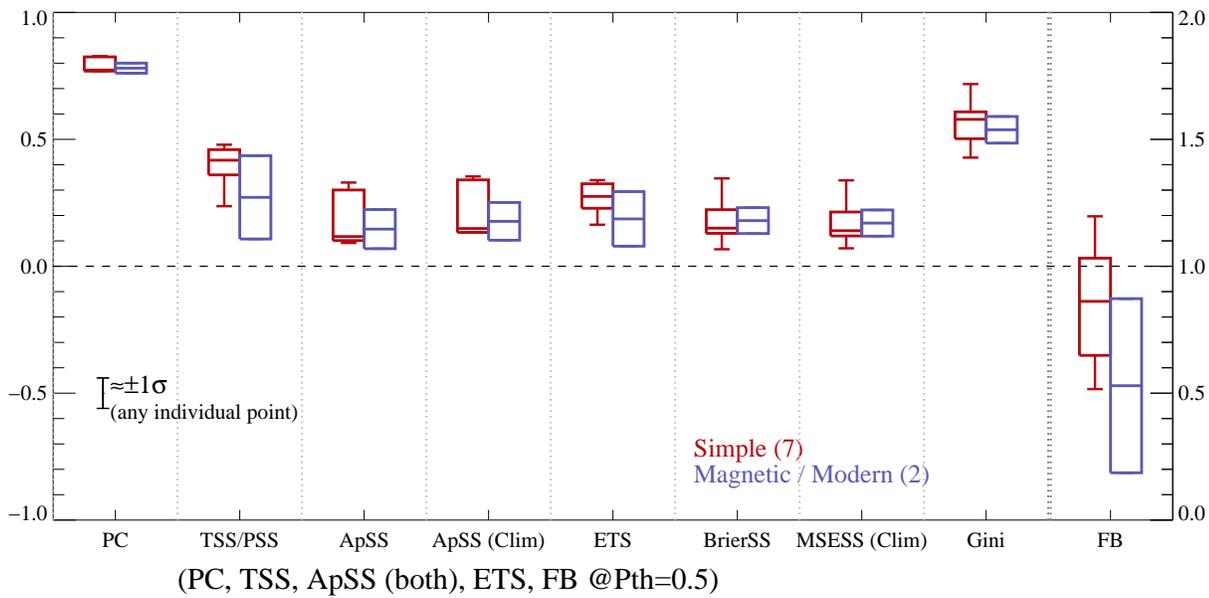}
\caption{Same as Figure~\ref{fig:comp_train}, but for comparisons of the parameters or data analysis used
by the forecasts, as indicated.}
\label{fig:comp_params}
\end{figure}

\begin{figure}[h]
\includegraphics[width=0.9\textwidth,clip, trim = 0mm 0mm 0mm 0mm, angle=0]{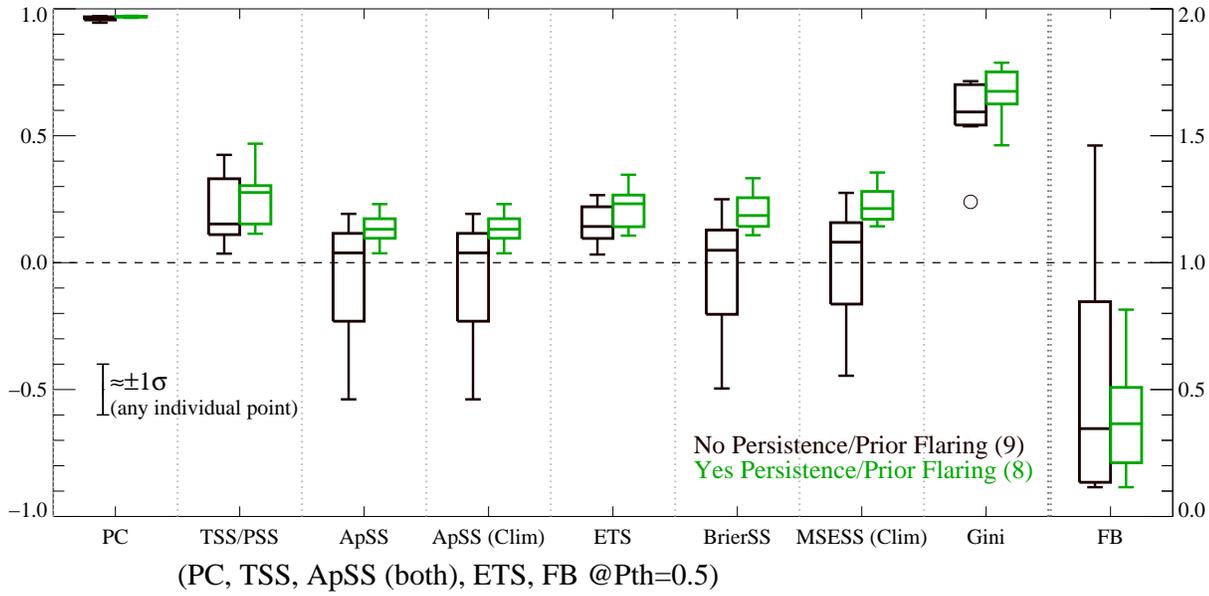}
\includegraphics[width=0.9\textwidth,clip, trim = 0mm 0mm 0mm 0mm, angle=0]{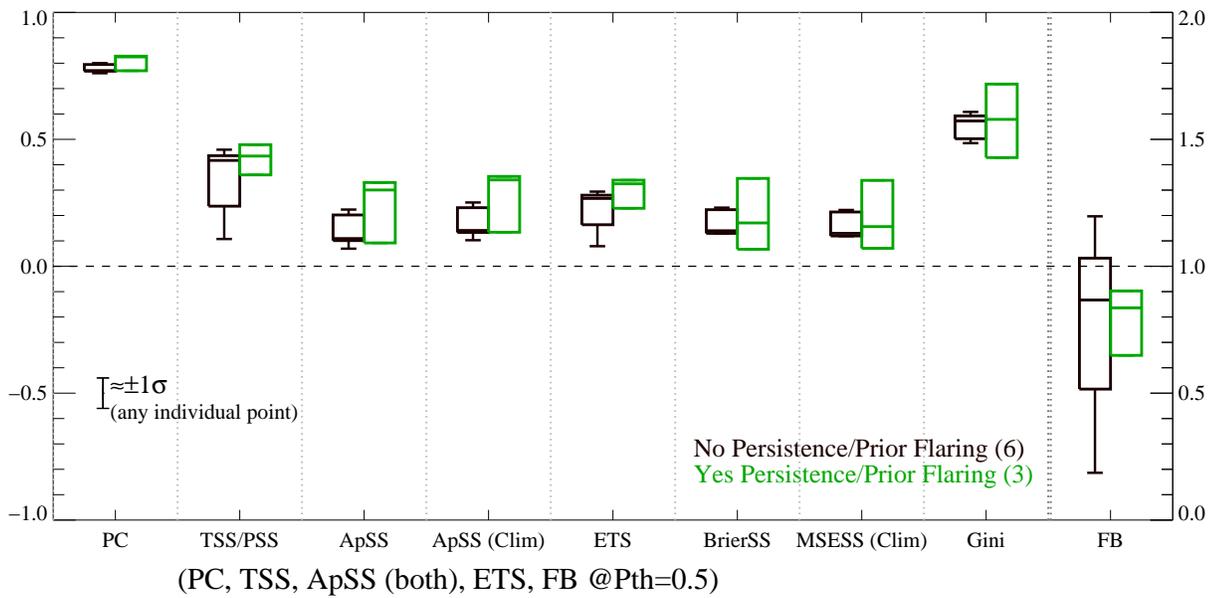}
\caption{Same as Figure~\ref{fig:comp_train}, but for the use of flare history or persistence
in the forecasts, as indicated.}
\label{fig:comp_persist}
\end{figure}

\begin{figure}
\includegraphics[width=0.9\textwidth,clip, trim = 0mm 0mm 0mm 0mm, angle=0]{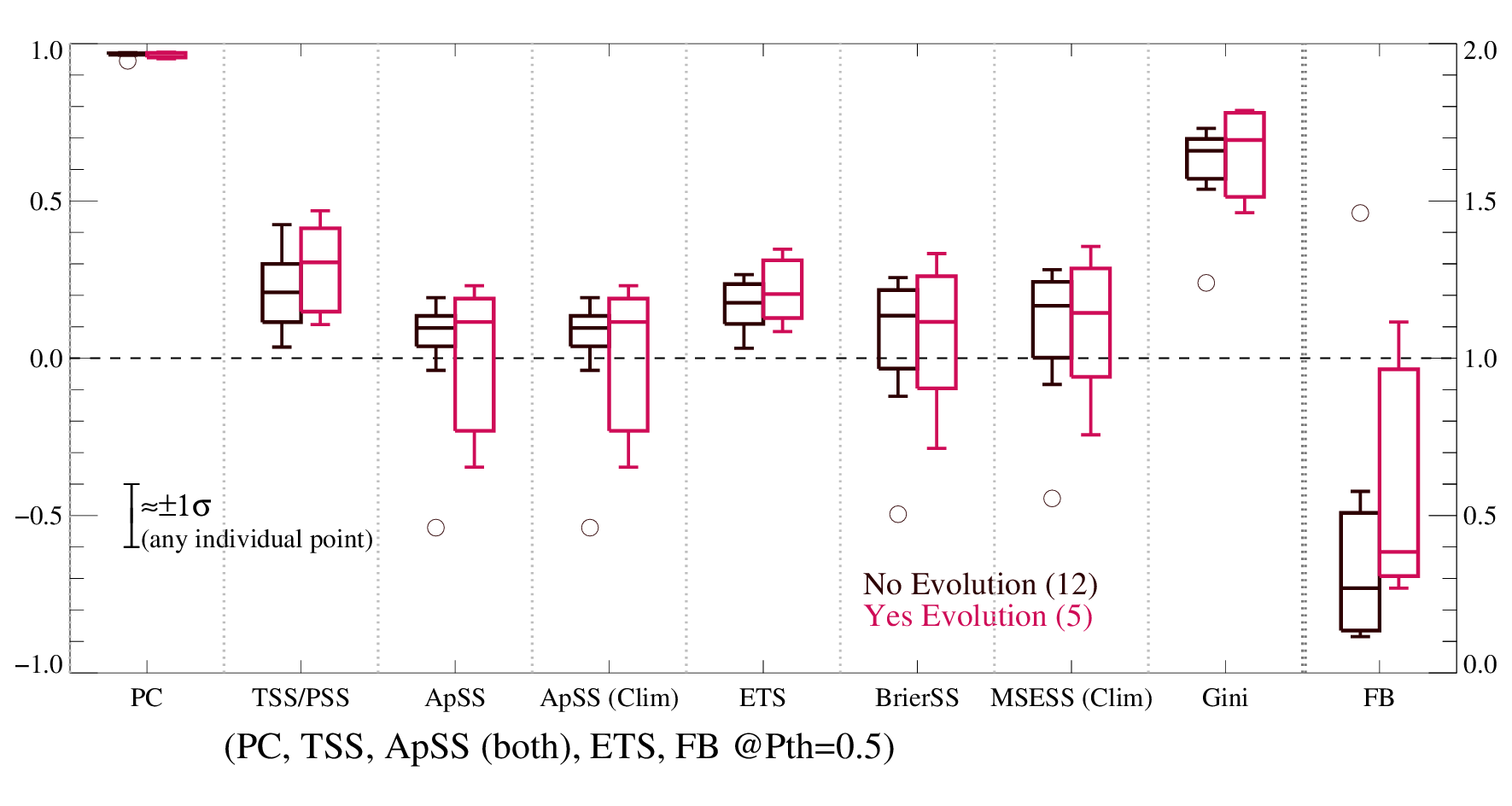}
\includegraphics[width=0.9\textwidth,clip, trim = 0mm 0mm 0mm 0mm, angle=0]{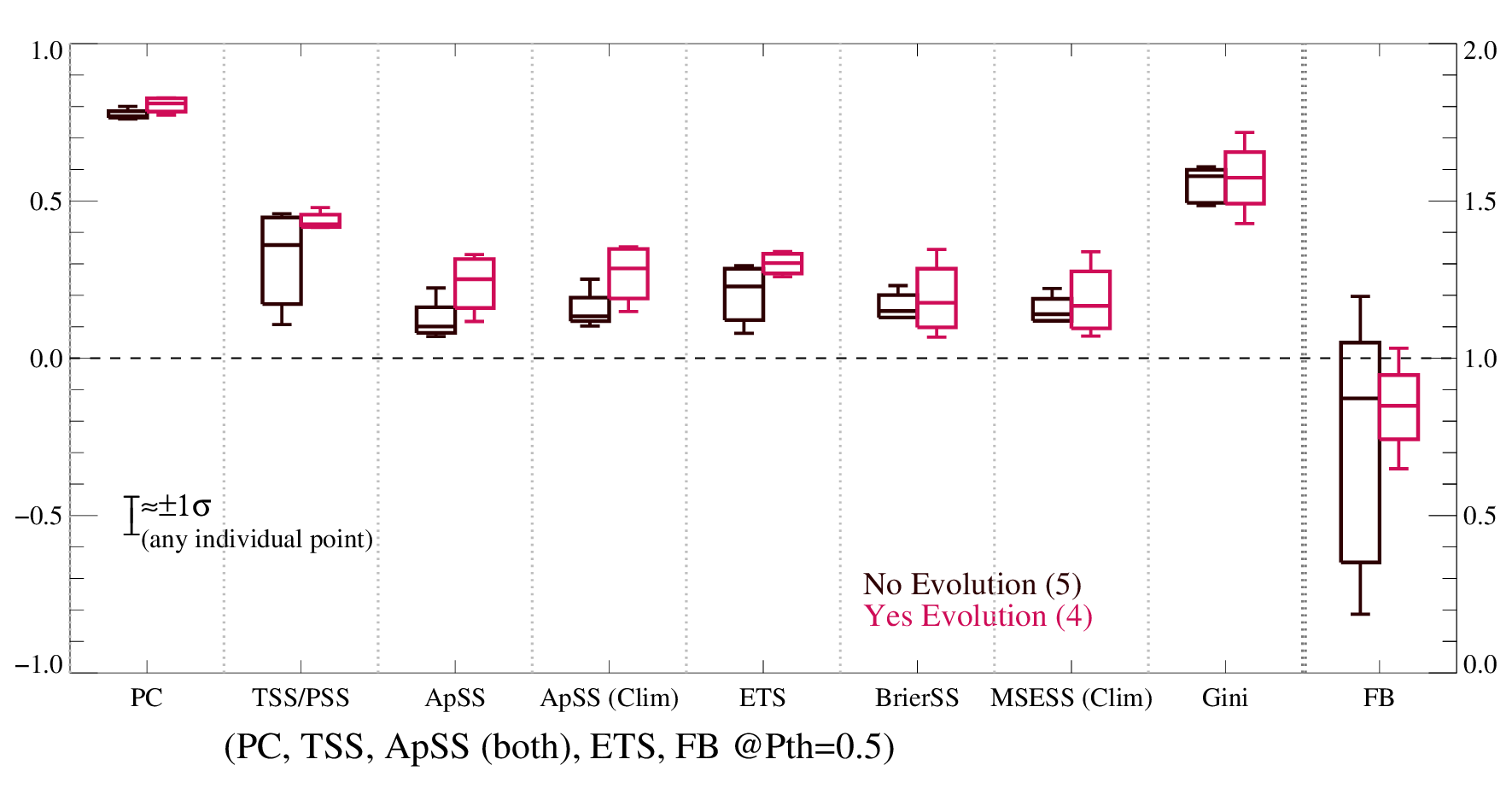}
\caption{Same as Figure~\ref{fig:comp_train}, but for the explicit use of active region evolution, as 
indicated.}
\label{fig:comp_evol}
\end{figure}

The results according to these categories are shown in
Figures~\ref{fig:comp_train}--\ref{fig:comp_evol}.  Throughout, the
estimated uncertainties in any one method's metric are of order $0.06$
for \CC\ and $0.10$ for \MM\ (see Paper II), are indicated on 
the box \& whisker plots, and should be kept in mind throughout
this discussion.  As discussed in Paper II, there is no single method
or group of methods that obviously out-perform the others.  There are
significantly fewer methods that produce \CC\ forecasts than produce \MM\
forecasts, but the event-category sample size is significantly smaller
for the latter, leading to larger estimated uncertainty in the metrics.

Generally speaking, the trends are not strong.  There is no trend present
which is present beyond the indicated quartiles across all metrics.
This is likely due to
a combination of factors including small sample size and significant
duplicity between method approaches, causing overlap between
different categories.  Additionally, as discussed above, there are
numerous subtleties whose influence cannot be 
captured in this analysis approach.
That being said, the trends are quite consistent across the metrics
(excluding FB and sometimes excluding PC).  The trends discussed here
are identified by means of weak but consistent (or dominant) trends in the
median score or the highest score as shown in the box \& whisker plots
({\it i.e.}, Figures~\ref{fig:comp_train} -- \ref{fig:comp_evol}).

From Figure~\ref{fig:comp_train} we see that ``short'' training intervals
(presumably on more modern / high-quality data) do not present any
obvious disadvantage (or any strong advantage).  The use of ``long'' training
intervals may be slightly disadvantageous for some metrics, in particular those
employing a climatological reference.  ``Long'' training also provided a much
wider range in FB to bring the range farther from ``significantly underforecasting''
results than the short or hybrid members.

The results in Figure~\ref{fig:comp_fcastmethod} indicate that at this point 
there is a slight advantage to using a statistical classifier (``ML/Classifier'') 
as compared to other correlations or Poisson statistics-based approaches (``Not ML'');
the trend is weak and only holds for a majority but not all of the metrics.
However, including a human (``Forecaster In The Loop''), does appear to
be systematically (albeit only slightly) advantageous.

From Figure~\ref{fig:comp_zones} there is a clear
disadvantage to using ``Restricted'' data for forecasts, compared to
``Full Disk'' forecasting.  For the \Mp\ event definition there is 
arguably a slight advantage to ``Earth-impacting'' forecasts over ``Full Disk''.

Figure~\ref{fig:comp_params} shows that there is a slight advantage 
according to climatology-referenced metrics to 
using ``Magnetic / Modern'' (quantitative) parameters for the
\MM\ tests.  However, there is a trend for better results according to FB and other
metrics for using ``Simple'' (qualitative) inputs or for the \CC\ event
definition.

Including persistence yields an improved performance across metrics
and event definitions, as evidenced
in Figure~\ref{fig:comp_persist}.  This may not be a surprise,
in that persistence has been a long-recognized indicator
of continued flare activity \citep{flareprediction,Bloomfield_etal_2016} and is 
often seen as the unofficial ``method to beat''.  A similarly
long-recognized indicator, the rapidity and character of evolution of the 
host active region, shows an advantage here in Figure~\ref{fig:comp_evol} as 
its inclusion provides better outcomes across at least a few metrics.

There are groups of methods which are similar enough across their
implementation that we may draw some interpretations.   In doing so
we refer to both the figures in this paper and the results and
figures in Paper II.  

First, the FITL methods were classified identically across our
characteristics groupings.  They generally employ similar tools
at the outset, those being long-trained historical flaring rates
following region classification according to size, complexity, {\it
etc.} \citep{McIntosh1990,flareprediction}.  Differences between
methods do arise through the additional tools -- both quantitative and
qualitative -- that are available at each center, but we did not
track those differences.   All FITL centers commonly have access to
(and fully utilize) a very wide selection of data sources; the humans
subjectively incorporate the presence of bright beyond-limb emission or
other indications of activity sources beyond the visible disk to extend
forecasts to beyond that from just the visible magnetic active regions.
The final input comes from humans.  Other studies have examined the
degree of influence that human input imparts to their facility's initial automated forecasts
\citep{Devos_etal_2014,Crown2012,Murray_etal_2017}.  The general trend between
those studies and here is consistent: human forecasters in the loop
add some skill.  Automated methods may be able to incorporate many of
these human-brought aspects to their forecasts in due time but, as of
yet, none do effectively.

Second, AMOS and MCEVOL are classified identically (MCSTAT differing only in the 
lack of incorporating evolution); morphologically their Reliability Plots
and ROC plots (Paper II Figures 2 and 3) appear similar.  While the MCEVOL
scores significantly worse on the climatology-referenced metrics 
than AMOS or MCSTAT ({\it i.e.}, the 
ApSS- and MSESS-based metrics), of interest here is that these
three are the only ``Long'' training-interval methods that do not
employ some other advancement such as machine learning, persistence,
or FITL.  The ``Long'' training-interval methods show
some detriment or longer negative-skill extents for some metrics.
In conjunction with the performance of the known members of the
group, this pattern leads to the conclusion that solely relying on historical flaring
rates (plus consideration for just active region growth) is insufficient
for successful forecasting.  An underlying reason may be the influence
of varying climatology, in that these three methods heavily rely upon
prior-cycle training when the climatological flaring rate was significantly
higher than during our testing period; additionally, MCSTAT and MCEVOL train using 
data from SC22 while AMOS does not.  Training during a period of higher climatology
and forecasting during a period of lower flaring rate 
can lead to overforecasting, and this situation may poignantly
demonstrate of the impact of variable climatology \citep{McCloskey_etal_2018}.

Two methods lie at the other end of automation, with DAFFS and BOM the sole
members of the ``Short''-training group.  Both rely primarily on 
high-quality ({\it SDO}/HMI) data and magnetic or modern parametrizations,
include measures of prior flaring, and employ ML/statistical classifier tools.
They tend to under-forecast according to the FB metric (DAFFS slightly less so,
see Paper II Figure 4), but perform similarly in other metrics (for the \MM\
grouping, since BOM does not provide \CC\ forecasts).  If one accounts
for the performance of the other members of the ``ML/Classifier'' group,
it strengthens the support for a conclusion that there is significant
overall skill brought by the combination of approaches illustrated 
by these two methods.

All FITL centers also have protocols (often some form of climatology)
for providing a default or ``fall-back'' forecast; there are no outtages.
This is a quality that some of the automated methods have invoked through
repeating the prior forecast, falling back to climatology or, in the
case of DAFFS, a progression to DAFFS-G, persistence-measures only, and
finally to climatology upon worsening data availability.  The performance
of methods that lack a default forecast is penalized by the evaluations
carried out here and, as discussed in Paper II, can be symptomatic of a
marked difference between the research and operational phases of a method.

\section{Discussion}
\label{sec:disc}

We examine the performance of operational flare-forecasting facilities over
a standardized testing interval and using standardized event definitions,
with the tools of quantitative evaluation metrics.  The limited number of
events over the testing interval plus the limited number of distinctly
different methods make it difficult to draw firm conclusions.  However,
upon examining the results according to particular implementation
techniques and details, a few trends emerge.

The strongest results show that, operationally, the long-held
``forecaster's wisdom'' of forecasting increased flare probability
from complex and evolving active regions that flared previously is
fairly successful.  In some cases there are methods that now put these
characteristics onto a quantitative basis, although for other methods
these aspects are still only incorporated qualitatively.  While there is still
a spread for some metrics and not fully consistent behavior across all
metrics, this appears to be a clear trend.

The use of modern data (such as from the {\it SDO}/HMI instrument)
or the quantitative analysis of magnetic field data appears to have no
significant effect on the performance, providing no obvious advantages
at this point but also providing no disadvantages.

Modern statistical methods are now employed in a number of ways for
operational forecasting.  A few methods have used machine-learning
techniques to identify and classify sunspot groups, others use
machine-learning algorithms and statistical classifiers to quantify
the parameter-space behavior of active regions.  Those methods in the
former category, however, then generally rely on a Poisson-statistics
based analysis of historical flare rates, while there are only three
methods that presently incorporate machine learning for the forecast
production itself.  As such, the sample sizes and limitations of this
comparison mean that we cannot comment on any advantages of machine
learning in operational flare forecasting.

That being said, the over-arching result of both Paper
II and the present study is that none of the current operational flare
forecasting methods perform exceptionally well across all performance
metrics.  However, we may begin to understand some reasons behind particularly
poor or particularly good performance in some cases.  

Most notably, this study is the first systematic demonstration of {\it
how} to engage in head-to-head comparisons of operational forecasting
models in order to recognize useful trends for future improvements
and development.  We extend this further in Paper IV \citep{ffc3_3} 
with a new method that focuses on temporal patterns
of forecasting errors.
Lessons learned from this community effort can
help guide future efforts to compare forecasts (such as
forecasts collected by the NASA/CCMC Flare Scoreboard\footnote{{\tt
https://ccmc.gsfc.nasa.gov/challenges/flare.php}}) and perhaps help
solidify the understanding of what approaches significantly improve
performance.

\appendix

\section{Participating Methods and Facilities}
\label{sec:method_table}

In Table~\ref{tbl:methods} we reproduce an abbreviated version of Table 1 
from Paper II, listing the methods and facilities involved with this work and the
monikers used to refer to them.

\begin{table}
\caption{Participating Operational Forecasting Methods (Alphabetical by Label Used)}
\label{tbl:methods}
%\advance\tabcolsep-4pt
\tabletypesize{\scriptsize}
\begin{center}
\begin{tabular}{p{6cm}p{5cm}p{2cm}p{4cm}}
Institution & Method/Code Name & Label & Reference(s) \\ \hline \hline
ESA/SSA A-EFFORT Service & Athens Effective Solar Flare Forcasting & A-EFFORT & \citet{GeorgoulisRust2007}\\ \hline
Korean Meteorological Administration \& Kyung Hee University & Automatic McIntosh-based Occurrence probability of Solar activity & AMOS & \citet{amos}\\ \hline
University of Bradford (UK) & Automated Solar Activity Prediction & ASAP & \citet{ColakQahwaji2008,ColakQahwaji2009} \\ \hline
Korean Space Weather Center (by SELab, Inc) & Automatic Solar Synoptic Analyzer & ASSA &\citet{ASSA}, \citet{ASSA_DOC} \\ \hline
Bureau of Meteorology (Australia) & FlarecastII & BOM  &\citet{Steward_etal_2011,Steward_etal_2017}\\ \hline
120-day No-Skill Forecast & Constructed from NOAA event lists & CLIM120  & \citet{SharpeMurray2017} \\ \hline
NorthWest Research Associates (US) & Discriminant Analysis Flare Forecasting System & DAFFS  &\citet{nci_daffs} \\ \hline
 '' '' & GONG+GOES only & DAFFS-G  &'' ''  \\ \hline
  NASA/Marshall Space Flight Center (US) & MAG4 (+according to & MAG4W & \citet{Falconer_etal_2011}; \\ 
  '' '' & magnetogram source  & MAG4WF & also see Paper II, Appendix A \\ 
  '' '' & and flare-history  & MAG4VW & \\ 
  '' '' & inclusion) & MAG4VWF & \\ \hline
Trinity College Dublin (Ireland) & SolarMonitor.org Flare Prediction System (FPS)  & MCSTAT  & \citet{gallagheretal02,Bloomfield_etal_2012}\\ \hline
  '' '' & FPS with evolutionary history & MCEVOL  & \citet{McCloskey_etal_2018} \\ \hline
MetOffice (UK)  &  Met Office Space Weather Operational Center human-edited forecasts & MOSWOC  & \citet{Murray_etal_2017}\\ \hline
National Institute of Information and Communications Technology (Japan) &  NICT-human  & NICT & \citet{Kubo_Den_Ishii_2017}\\ \hline
New Jersey Institute of Technology (UK) & NJIT-helicity & NJIT  & \citet{Park_Chae_Wang_2010}\\ \hline
NOAA/Space Weather Prediction Center (US) & & NOAA & \citet{Crown2012}\\ \hline
Royal Observatory Belgium Regional Warning Center & Solar Influences Data Analysis Center human-generated & SIDC & \citet{Berghmans_etal_2005,Devos_etal_2014}\\ \hline
\end{tabular}
%\footnotetext{$\dagger$: if applicable}
\end{center}
\end{table}

\section{Acronyms}
\label{sec:tlas}

Acronyms and references used in Tables~\ref{tbl:questions1}--\ref{tbl:questions4} 
are expanded upon here.

\begin{list}{}{
\setlength{\topsep}{0cm}   % vertical space between text, 1st line of list
\setlength{\partopsep}{0cm} %vertical space "  "  if preceded by blank line
\setlength{\itemsep}{0cm}   % space between items
\setlength{\parsep}{0cm}    % vertical space between Paragraphs w/in item
\setlength{\leftmargin}{0cm} % left margin -> left margin of list
\setlength{\rightmargin}{0cm}% right margin of list -> right margin
\setlength{\listparindent}{0cm} % extra indentation to paragraphs w/in list
\setlength{\itemindent}{0cm} % indentation of first line of items
\setlength{\labelsep}{0cm}   % space between label and text of item
\setlength{\labelwidth}{0cm} % width of box containing label, > 0
%\makelabel{}
%\usecounter{}
}

\item {\bf AIA}: Atmospheric Imaging Assembly (on {\it SDO}) \citep{aia}
\item {\bf ApSS:} Appleman Skill Score
\item {\bf AR:} Active Region
\item {\bf BrierSS:} Brier Skill Score
\item {\bf CM:} Central Meridian
\item {\bf ETS:} Equitable Threat Score
\item {\bf EUVI:} Extreme Ultraviolet Imager (on {\it STEREO}) \citep{euvi}
\item {\bf FB:} Frequency Bias
\item {\bf FD:} Full Disk
\item {\bf GOES:}  Geostationary Observing Earth Satellite (run by NOAA)
\item {\bf GONG:} Global Oscillations Network Group \citep{gong}
\item {\bf HARP:} HMI Active Region Patch \citep{hmi_pipe,hmi_sharps}
\item {\bf HMI:} Helioseismic and Magnetic Imager \citep{hmi_pipe}
\item {\bf MSESS:} Mean Square Error Skill Score
\item {\bf NRT:} Near Real Time (data)
\item {\bf PC:} Proportion Correct (also known as Rate Correct)
\item {\bf PDE:} Probability Density Estimate
\item {\bf PROBA2/SWAP:} PRoject for Onboard Autonomy / Sun Watcher using Active Pixel System detector and Image Processing
\item {\bf SC:} Solar Cycle
\item {\bf SDO:} Solar Dynamics Observatory \citep{sdo}
\item {\bf SHARP parameters:} pre-computed ``Space Weather HARP'' parameters describing 
the magnetic field of HARP regions ({\it e.g.} total unsigned magnetic flux, 
total unsigned vertical current, {\it etc.} \citep{hmi_sharps}
\item {\bf SOON:} Solar Optical Observing Network
\item {\bf SRS:} Solar Region Summary, data product of NOAA/SWPC listing active-region 
attributes\footnote{Available from {\tt https://www.swpc.noaa.gov/products/solar-region-summary}.}.
\item {\bf STEREO:} Solar TErrestrial RElations Observatory \citep{stereo}
\item {\bf TSS:} True Skill Statistic (also known by Peirce Skill Score `PSS',
 Hanssen \& Kuiper Skill Score `H\&KSS', 
\item {\bf USAF:} US Air Force
\item {\bf Zpc:} Modified Zurich Classifications of sunspot groups

\end{list}

\acknowledgments

We wish to acknowledge funding from the Institute for Space-Earth
Environmental Research, Nagoya University for supporting the workshop and
its participants.   We would also like to acknowledge the ``big picture''
perspective brought by Dr. M. Leila Mays during her participation in
the workshop.  K.D.L. and G.B. acknowledge that the DAFFS and DAFFS-G tools
were developed under NOAA SBIR contracts WC-133R-13-CN-0079 (Phase-I) and
WC-133R-14-CN-0103 (Phase-II) with additional support from Lockheed-Martin
Space Systems contract \#4103056734 for Solar-B FPP Phase E support.
A.E.McC. was supported by an Irish Research Council Government of Ireland
Postgraduate Scholarship. D.S.B. and M.K.G were supported by the European Union
Horizon 2020 research and innovation programme under grant agreement
No. 640216 (FLARECAST project; {\tt http://flarecast.eu}).  MKG also 
acknowledges research performed under the A-EFFort project and subsequent 
service implementation, supported
under ESA Contract number 4000111994/14/D/ MPR.  S. A. M. is
supported by the Irish Research Council Postdoctoral Fellowship Programme
and the US Air Force Office of Scientific Research award FA9550-17-1-039.
The operational Space Weather services of ROB/SIDC are partially funded
through the STCE, a collaborative framework funded by the Belgian Science
Policy Office.

%\bibliography{/export/home/leka/TeX/kdl_biblio}

\end{document}